\documentclass[11pt,twocolumn,twoside,a4paper,amsmath,amssymb,aps,showkeys,showpacs]{revtex4}

\usepackage{amsfonts}
\usepackage{graphics} 
\usepackage{epsfig}
\usepackage{fancyheadings}

\begin{document}
\thispagestyle{myheadings}
\rhead[]{}
\lhead[]{}
\chead[SVD-2 Collaboration]{NEUTRAL PION NUMBER FLUCTUATIONS AT HIGH MULTIPLICITY 
IN pp-INTERACTIONS AT 50 GeV}

\title{NEUTRAL PION NUMBER FLUCTUATIONS AT HIGH MULTIPLICITY 
IN pp-INTERACTIONS AT 50 GeV}

\author{
{\bf (SVD-2 Collaboration)}\\
A. G. Afonin, E. N. Ardashev, V. F. Golovkin, S. N. Golovnya, S. A. Gorokhov, A. A. Kiryakov, A. G. Kholodenko, V. V. Konstantinov, L. L. Kurchaninov, I. S. Lobanov, E. V. Lobanova, G. A. Mitrofanov, V. S. Petrov, A. V. Pleskach, M. K. Polkovnikov, V. N. Riadovikov*, V. N. Ronzhin, V. A. Senko,N. A. Shalanda, M. M. Soldatov, Yu. P. Tsyupa, A. P. Vorobiev, V. I. Yakimchuk, and V. N. Zapolskii}
\email{riadovikov@ihep.ru}
\affiliation{%
IHEP, Protvino, Moscow region, Russia.}%
\author{A. N. Aleev, V. V. Avdeichikov, V. P. Balandin, Yu. T. Borzunov, V. A. Budilov, Yu. A. Chentsov, N. F. Furmanets, G. D. Kekelidze, V. I. Kireev, E. S. Kokoulina, A. K. Kulikov, E. A. Kuraev, N. A. Kuzmin, \fbox{G. I. Lanschikov}, V. N. Lysan, V. V. Myalkovskiy, V. A. Nikitin, V. D. Peshehonov, Y. P. Petukhov, I. A. Rufanov, V. I. Spiryakin, A. V. Terletskiy, A. I. Yukaev, and N. K. Zhidkov}
\affiliation{%
JINR, Dubna, Moscow region, Russia.}%
\author{S. G. Basiladze, S. F. Berezhnev, G. A. Bogdanova, N. I. Grishin, Ya. V. Grishkevich, G. G. Ermakov, \fbox{P. F. Ermolov}, D. E. Karmanov, V. N. Kramarenko, A. V. Kubarovsky, A. K. Leﬂat, S. I. Lutov, M. M. Merkin, V. Popov, L. A. Tikhonova, A. M. Vishnevskaya, V. Yu. Volkov, A. G. Voronin, S. A. Zotkin, and E. G. Zverev}
\affiliation{%
SINP MSU, Moscow, Russia.}%
\author{M. A. Batouritski}
\affiliation{%
NC PHEP BSU, Minsk, Belarus.}%
\author{A. V. Karpov and A. Ya. Kutov}
\affiliation{%
Department of Mathematics Komi SC UrD RAS, Syktyvkar, Russia.}%
\begin{abstract}
The results of pion fluctuation measurements in SERP-E-190 experiment (project Thermalization) with 50 GeV proton beam irradiation of the liquid hydrogen target at SVD-2 setup are presented. The photons are detected in the electromagnetic calorimeter. MC modeling of photon detection has shown the linear dependence between number of photons in the calorimeter and the average number of neutral pions. Neutral pion number $N_0$ distributions for each total number of particles in an event $N_{tot}=N_{ch}+N_0$ are obtained after making corrections on the setup acceptance, triggering and efficiency of the event reconstruction. The scaled variance of neutral pion fluctuations, $\omega=D/<N_0>$, is measured. The fluctuations increase at $N_{tot}>$22. According to quantum statistics models it may indicate for the approaching to pion condensate conditions for high pion multiplicity in the system. This effect have been observed for the first time. 

\end{abstract}

\pacs{ 13.75.Cs, 24.60.Ky, 67.85.Hj, 29.85.Fj }

\keywords{ pp-interactions, pion fluctuation, pion condensate, simulation, data handling }

\maketitle


\renewcommand{\thefootnote}{\roman{footnote}}


\section{Introduction}
\label{introduction}

The SERP-E-190 experiment (project Thermalization) is carried out at upgraded setup SVD-2 (Spectrometer with Vertex Detector), which is irradiated with 50 GeV proton beam from U-70 IHEP (Protvino) accelerator. The setup is equipped with a liquid hydrogen target, a microstrip silicon vertex detector (VD), a drift tube tracker, magnetic spectrometer, Cherenkov counter and an electromagnetic calorimeter for photon detection (DEGA) \cite{Proposal}. Project Thermalization is aimed at studying processes with high multiplicity in pp-interactions, which are one of fundamental regions of hadron physics. It is not possible to describe them with perturbative QCD. The theory gives only a qualitative picture of these processes. The multiplicity distribution at 50 GeV was measured earlier \cite{Ammosov01} up to the number of charged particles $N_{ch}$=16 with average number of charged particles at this energy $<N_{ch}>$=5.3. The kinematical limit for the total number charged and neutral particles is $N_{tot}$=56. The data for $N_{ch}=4\div$22 and for $N_{tot}=4\div$31 are presented in this work. Collective effects can occur in the events with the multiplicity by several times higher than the average one: the large fluctuations of charged and neutral pion numbers, as result of formation pion condensate, formation of jets with identical pions, the so-called Bose-Einstein multiparticle effect, formation of ring events as result gluon hadronization, which are emitted out by partons in the nuclear environment (Cherenkov radiation analog) and others. The SVD-2 data allows one to check and develop various models of multiparticle production with $N_{tot}><N_{tot}>$.

M. I. Gorenstein and V. V. Begun \cite{Begun01} \cite{Begun02} have shown that at the approach of the pion system to Bose-Einstein condensate conditions (BEC) the neutral pion number fluctuations are increasing in accordance with the model based on quantum statistics. These fluctuations can be detected by the scaled variance, $\omega$, which is defined as the ratio of variance D for neutral pion number distribution $N_{0}$ and average $<N_{0}>$,
\begin{center}
$\omega = D/<N_0>$.
\end{center}
The value of $\omega$ rising with increasing of the total particle number, $N_{tot}=N_{ch}+N_0$, depends on temperature and energy density of pion system.

A part of E-190 data is used in this paper to search for pion number fluctuations. Numbers of charged particles, $N_{ch}$, and photons, $N_{\gamma}$, in each event are detected. These values are corrected on the setup acceptance and reconstruction efficiency by means of modeling. Simulation has allowed getting the neutral pion number, $N_0$, in each event also. Numbers of events, $N_{ev}(N_0, N_{tot})$, are measured here with corrections on various losses. For the analysis of the data at different $N_{tot}$ relative values of $n_0=N_0/N_{tot}$ and $r_0=N_{ev}(N_0, N_{tot})/N_{ev}(N_{tot})$ are used. Thus $n_0$ is changed in the range of 0$\div$1 and the sum of all $r_0$ is equal to 1 for each $N_{tot}$ (normalization condition). The Fig. 1 qualitatively illustrates the behaviour of value $r_0$ for three cases: a) the generation of events with PYTHIA5.6 program, b) the pion system in which some pions drop out into condensate, c) all pions are in the BEC condition. Each distribution is characterized by average, $<n_0>$, and by standard deviation, $\sigma$ for a Gaussian fit.
\begin{figure}
\includegraphics[scale=0.33]{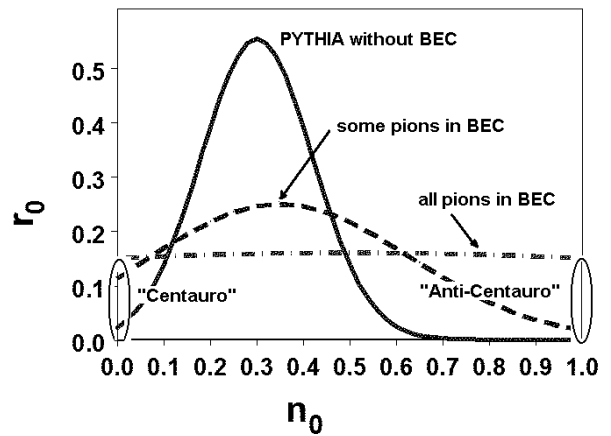}
\caption{\label{fig:wide}Distributions $r_0$ for normalized multiplicity of neutral pions in QCD model and when the system approaches to BEC.}
\end{figure}

\section{Simulation of neutral pion detection}
\label{simulation}

Calorimeter DEGA at SVD-2 setup detects the events with photons from neutral pions decay. Registration of all $\pi^0$ in the event is not possible because of limited DEGA aperture and the threshold on photon detection energy. But $\pi^0$ reconstruction efficiency can be estimated by means of simulation. Using PYTHIA5.6 code $10^6$ events (MC) are simulated for pp$\rightarrow$X inelastic interactions at 50 GeV. Only the events with $N_{ch}\geq$4 are analyzed. The photon detection efficiency is assumed to be equal to 1 if photon hits DEGA and its energy is greater than 100 MeV. Practically all photons are the product of $\pi^0$ decays (95\%), 37\% of them  give two gammas signal in DEGA and 18\% of $\pi^0$ result in one photon signal. Fig. 2 shows the dependence of pion number in event, $N_0$, on number of photons in DEGA, $N_{\gamma}$. It is clear that there is no unique connection between $N_{\gamma}$ and $N_0$. (Fig. 2a). Instead each $N_{\gamma}$ is associated with some number of $N_0$ and there is a linear correlation between average $<N_0>$ and $N_{\gamma}$ (Fig. 2b). So relation between the number of events 
$N_{ev}(N_{\gamma}, N_{ch})$ and $N_{ev}(N_0, N_{ch})$ can be found from this analysis. In Fig. 2c multiplicity distributions for pions and photons in DEGA are presented. The form of these distributions is similar except the area of small multiplicity.

\begin{figure}
\includegraphics[scale=0.4]{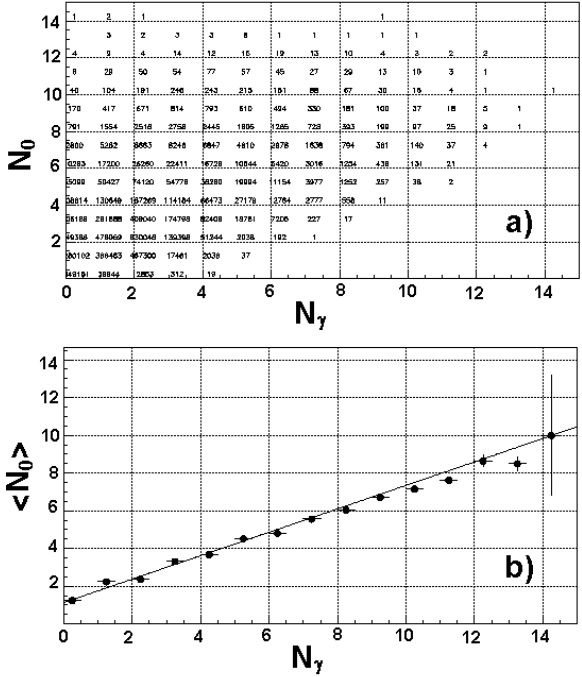}
\caption{\label{fig:wide}a) The dependence of the pions number in MC events on photons number detected in DEGA, \\
b) $<N_0>$ and $N_{\gamma}$ correlation }
\end{figure}

Values n and r are calculated for MC events. Then the dependence of r versus n has been fitted with Gaussian. In Fig. 3a the fitted parameters $<n_0>$ and $<n_{\gamma}>$ on $N_{tot}$ are presented, where $N_{tot}=N_{ch}+N_0$ for $\pi^0$ and $N_{tot}=N_{ch}+N_{\gamma}$ for photons. Standard deviations, $\sigma$ of Gaussians, are 
presented in Fig. 3b too. The normalized dispersion, $\omega$=$\sigma^2\ast$ $N_{tot}/<n>$ is shown in Fig. 3c. The value $\omega$ decreases for photons, but remains near the constant for pions in the full area of $N_{tot}$ changes.

\begin{figure*}
\includegraphics[scale=0.3]{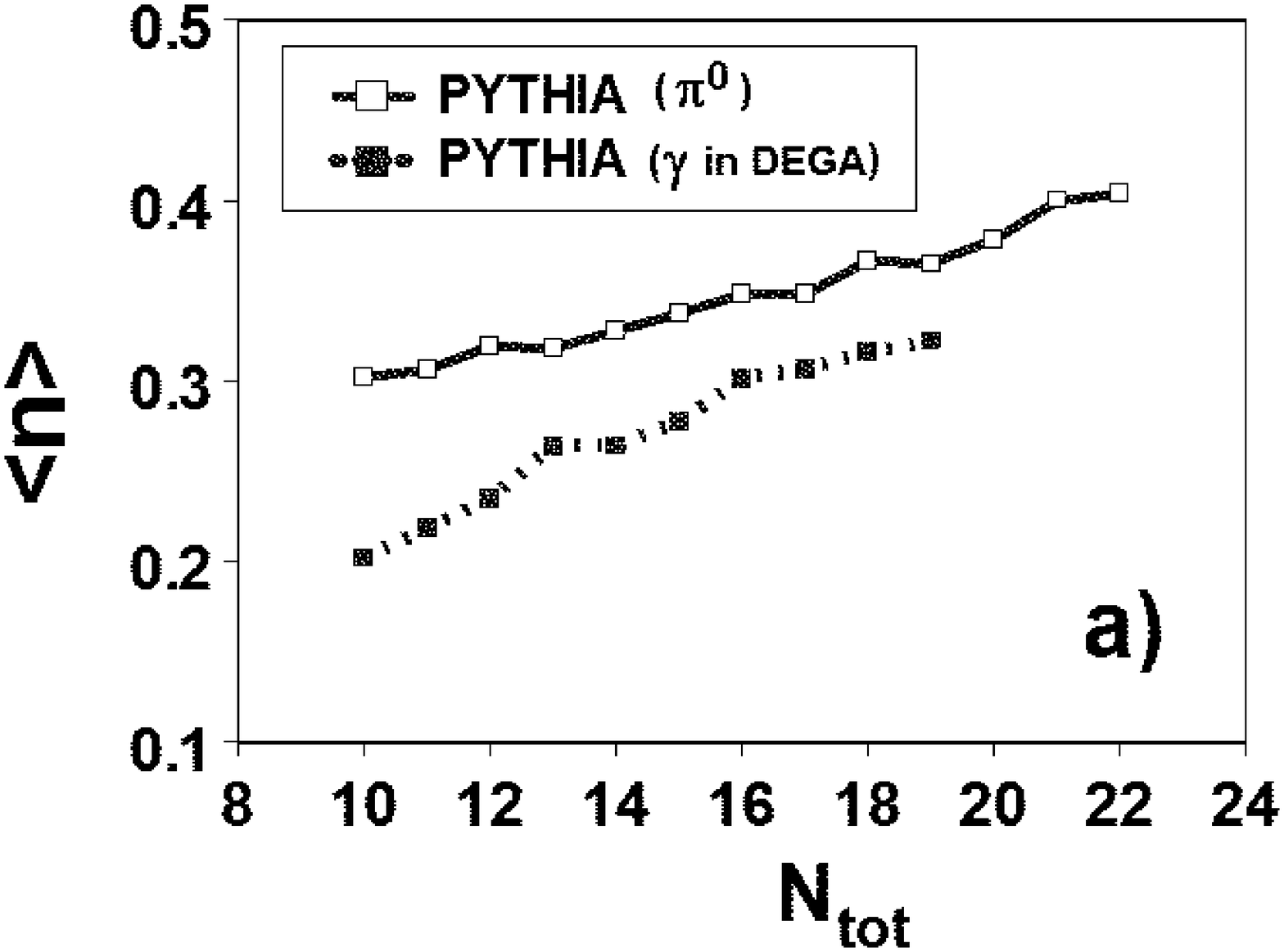}
\includegraphics[scale=0.3]{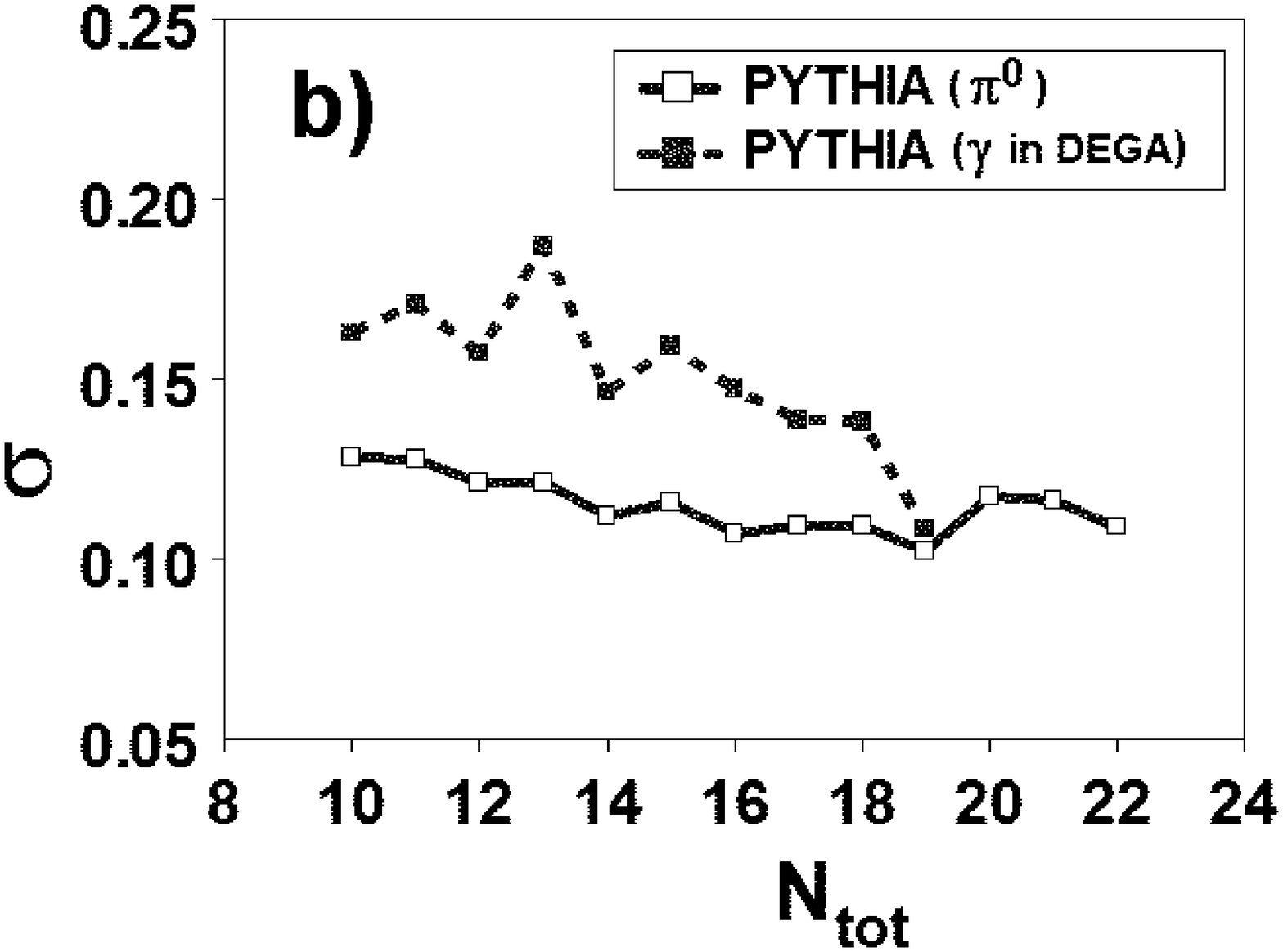}
\end{figure*}
\begin{figure*}
\includegraphics[scale=0.3]{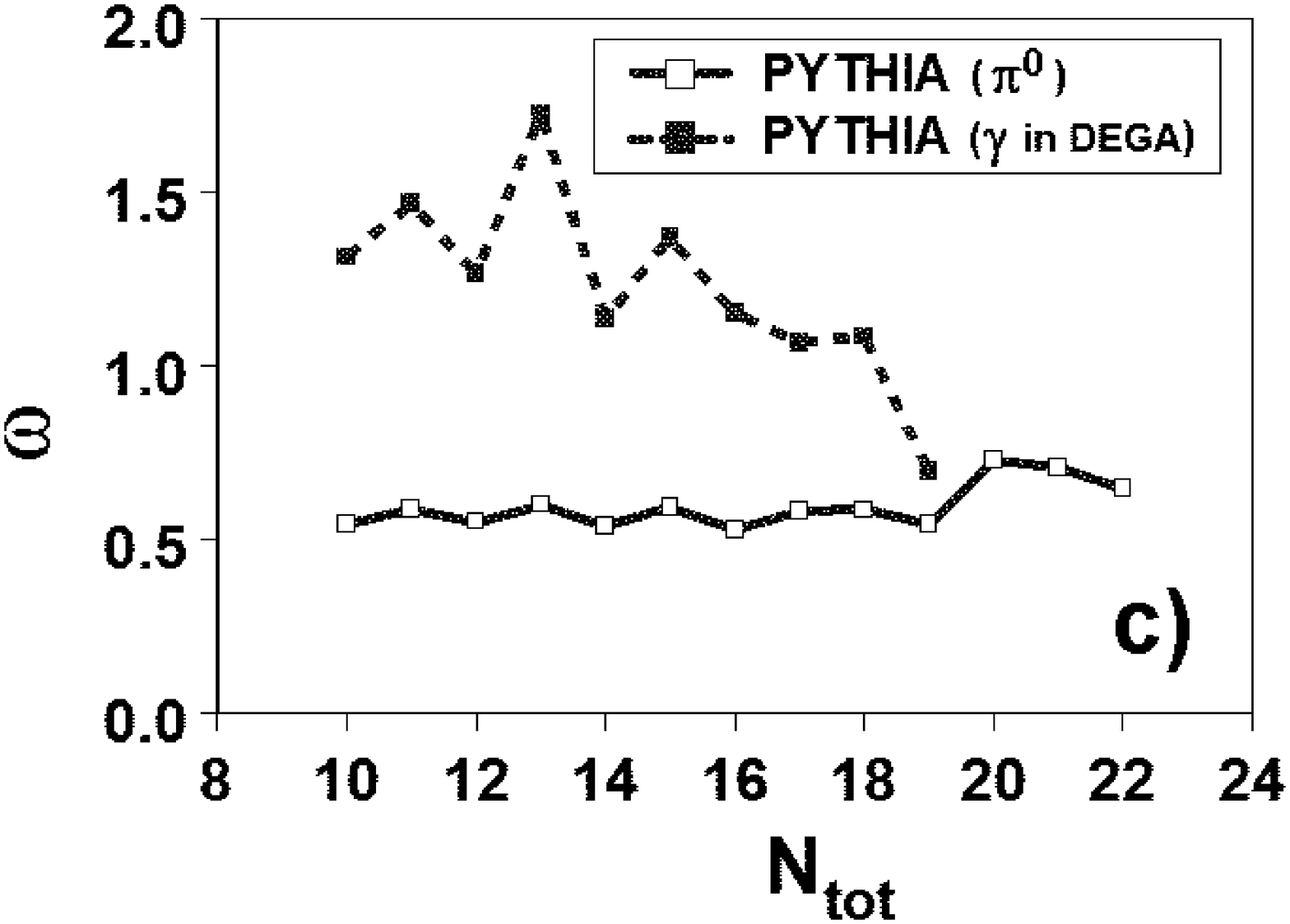}
\caption{\label{fig:wide}а) Average $<n_0>$, $<n_{\gamma}>$; b) standard deviation $\sigma$ and c) scaled variance $\omega$ dependence on $N_{tot}$ (see text) for MC events. .}
\end{figure*}

\section{Photon and charged particles reconstruction}
\label{reconstruction}

The calorimeter DEGA contains 42x32=1344 elements from lead glass blocks with PM. The calibration of DEGA is carried out with irradiation of each element center by 15 GeV electron beam. Almost all (98\%) energy of electromagnetic shower from the photon is located in the cell (3$\times$3) elements. Photon reconstruction consists in the searching of (3$\times$3) signal clusters and analyzing them with criteria for photon. The present work comprises $\sim5\ast10^5$ events of pp interaction.  The analysis of the electromagnetic showers has led to the following results: average photon number in the event is $<N_{\gamma}>$=3.0 (Fig. 4а), their average energy is $<E_{\gamma}>$=2.8 GeV (Fig. 5b), the minimum energy of photon detection is equal to 100 MeV.

For the charged tracks reconstruction the data only from VD have been used. Because of various losses the corrections of charged particle number are essential for measurements of pion fluctuations. The correction for the setup acceptance and the particle reconstruction efficiency is made by means of the apparatus performance simulation. The obtained weights have shown the contribution of events with a different true number of charged tracks into the event sample with the reconstructed multiplicity. The details of this procedure are presented in \cite{Ardashev01}.In the present work the number of events with measured $N_{ch}$ is distributed on event samples with restored value $N_{ch}$ according to these weights.

\begin{figure*}
\includegraphics[scale=0.3]{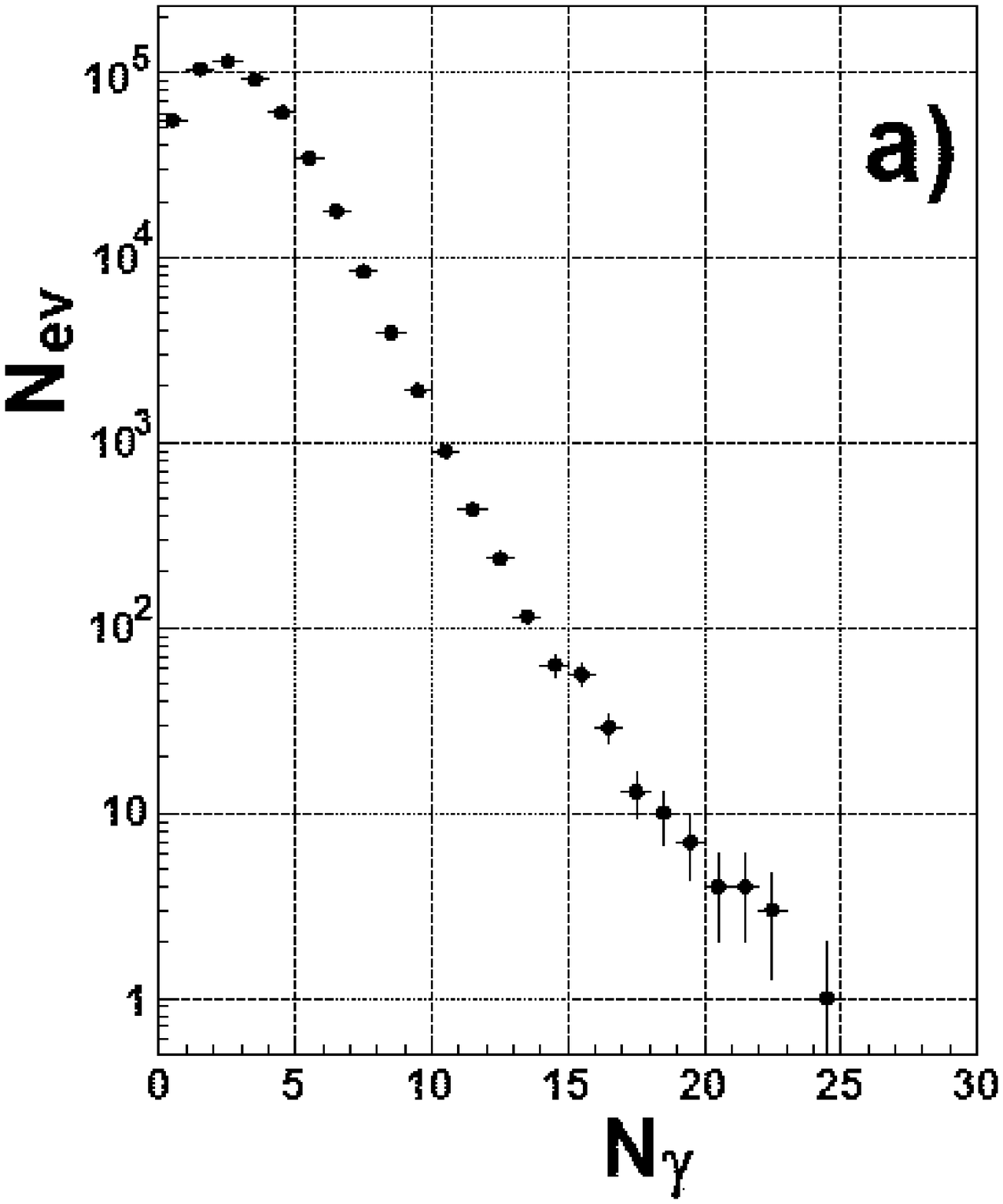}
\includegraphics[scale=0.3]{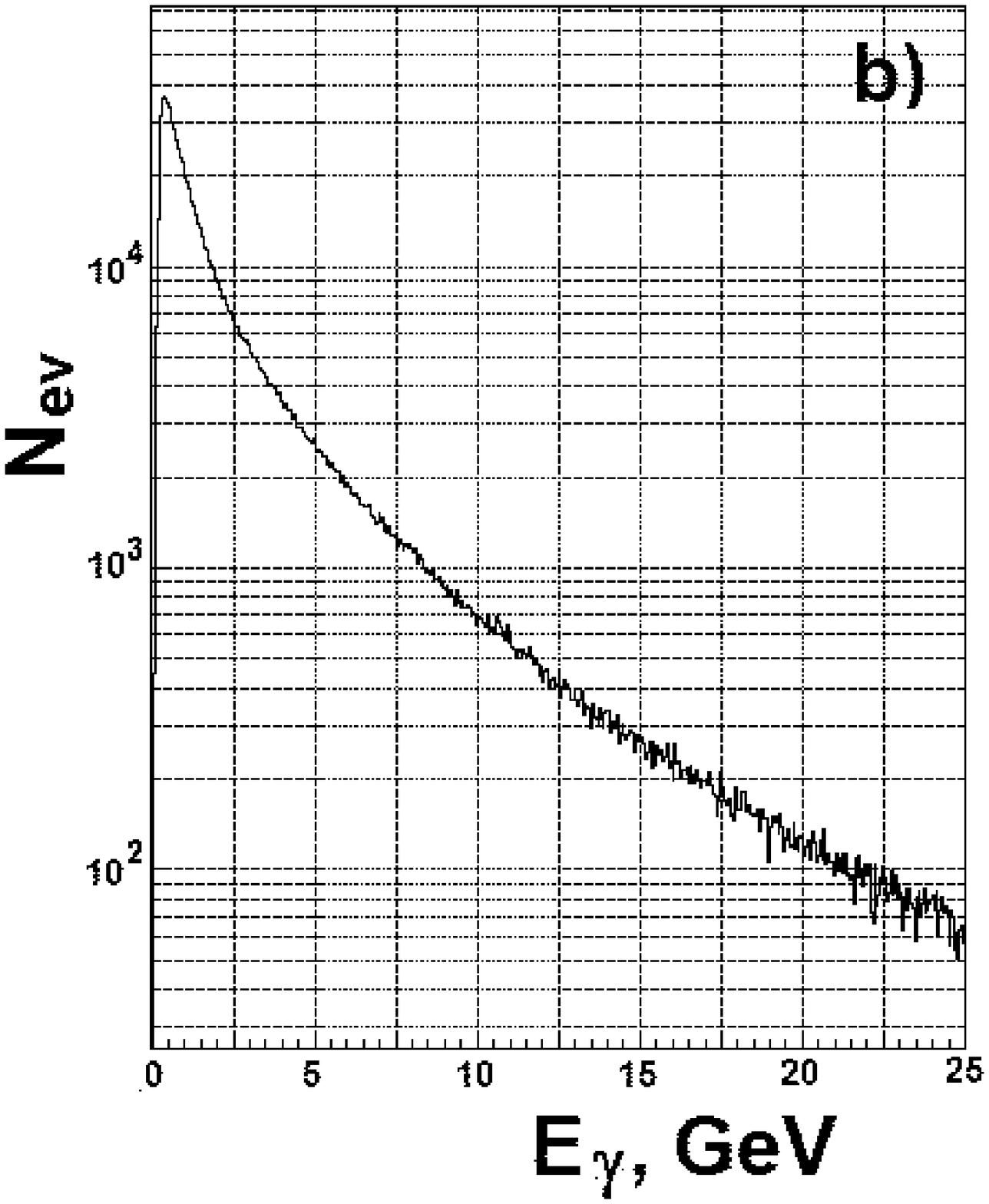}
\caption{\label{fig:wide}a) Photon multiplicity distribution; b) photon energy distribution for MC events.  }
\end{figure*}

Topological cross sections for pp-interactions at 50 GeV were measured in experiment with Mirabelle bubble chamber \cite{Ammosov01}. Taking into account that data topological cross sections for $10\leq N_{ch}\leq 24$ had been obtained in \cite{Ardashev01}. The present data are obtained with suppression of small charged multiplicity events ($N_{ch}<8$) with trigger conditions. Therefore the charged multiplicity distribution has been corrected to lead to the distribution received in \cite{Ammosov01} \cite{Ardashev01}.

\begin{figure}
\includegraphics[scale=0.35]{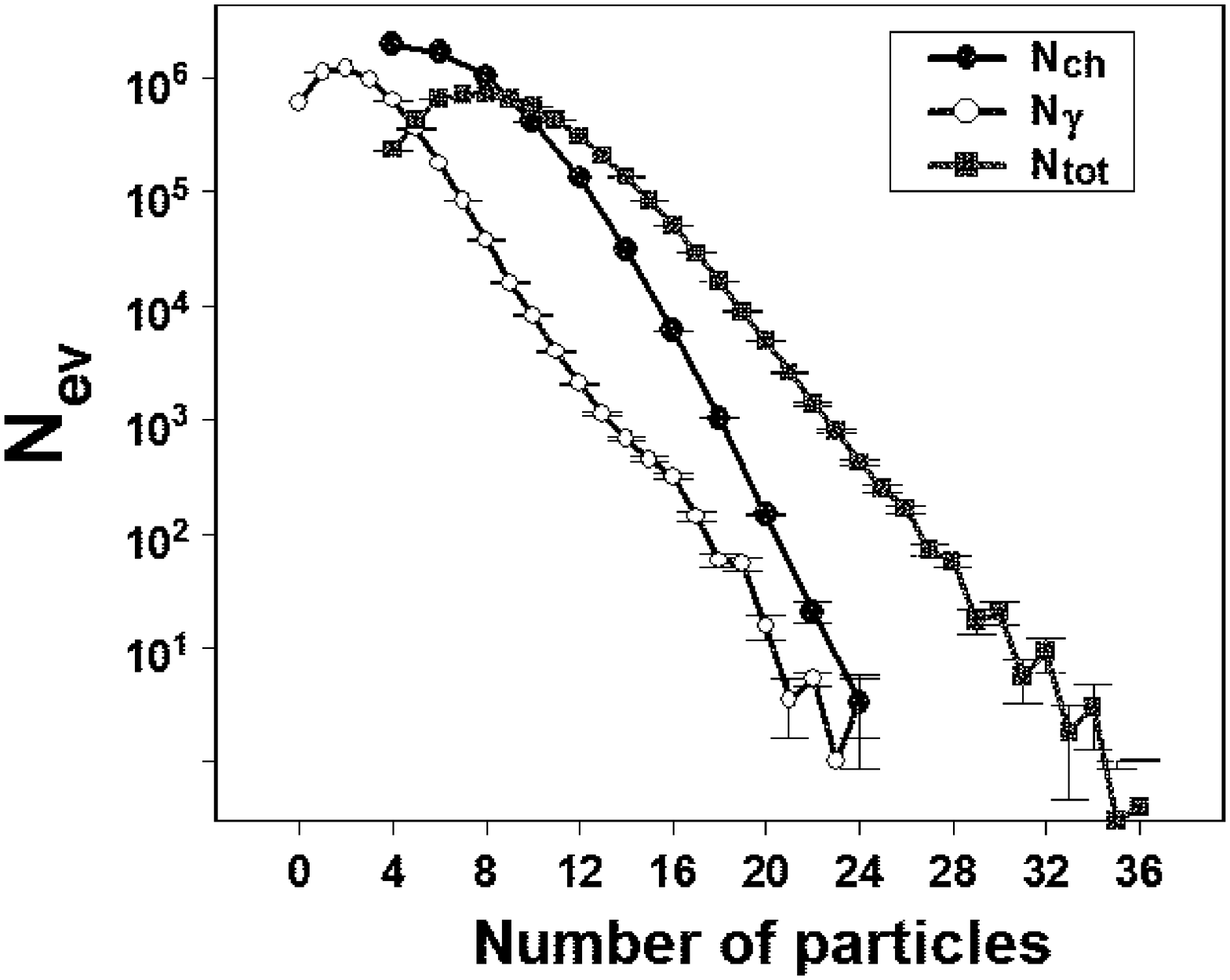}
\caption{\label{fig:wide}Distributions of corrected $N_{ch}$, $N_{\gamma}$ and $N_{tot}=N_{ch}+N_{\gamma}$.}
\end{figure}

Fig. 5 shows charged multiplicity distributions after corrections. It is necessary to stress that the change of the event number for $N_{ch}$ after introduction of corrections also leads to the change of the event number for $N_{\gamma}$. So, the distribution for $N_{\gamma}$ in Fig. 5 differs from the similar distribution in Fig. 4a for observable event numbers. 

\section{Neutral pion fluctuation measurements}
\label{fluctuation}

Thus we have corrected event numbers $N_{ev}(N_{\gamma}, N_{ch})$. Then two-dimensional $N_{ev}(N_{\gamma}, N_0)$ distributions for MC events (see Fig. 2а) are used to recover event numbers $N_{ev}(N_0, N_{ch})$. We have introduced notations i=$N_{\gamma}$, j=$N_0$ and $N_{ev}(N_{\gamma}, N_0)=N_{ev}(i,j)$. For each $N_{ch}$ matrix of coefficients $c_{ij}=N_{ev}(i,j)/N_{ev}(i)$ is calculated, where $N_{ev}(i)=\sum_jN_{ev}(i,j)$. Event number $N_{ev}(N_{\gamma}, N_{ch})$ is decomposed in sums of events with various $N_0$, $N_{ev}(i,j)=c_{ij}\ast N_{ev}(i)$ at $N_{ch}$=const. Normalization condition $\sum c_{ij}$=1 is satisfied. Resulting sum $N_{ev}(j)={\sum_i}N_{ev}(i,j)$ at $N_{ch}$=fix is the analog of event number $N_{ev}(N_{\gamma}, N_{ch})$, but for pions.

The simulation by PYTHIA5.6 allows to obtain $c_{ij}$ for $N_{\gamma}\leq10$ and $N_{ch}\leq14$ only because of limitation of the MC events statistics. Regularities of factors $c_{ij}$ are used to continue them to $N_{\gamma}\geq10$ and $N_{ch}\geq14$ region. Fig. 6a illustrates the dependence of $c_{ij}$ factors on $N_0$ for various $N_{\gamma}$ and $N_{ch}$. The form of these distributions slightly depends on $N_{\gamma}$ and $N_{ch}$, but their average $<N_0>$ increases with $N_{\gamma}$. The dependence of the average $<N_0>$ and standard deviation (rms) on $N_{\gamma}$ are shown in Fig. 6b. After fitting it by linear dependence the coefficients $c_{ij}$ for $N_{\gamma}\geq$10 and $N_{ch}>$14 are calculated and the full sample of $N_{ev}(N_{tot}, N_{ch}, N_0)$ is obtained, which is used then to determine pion fluctuation.

\begin{figure*}
\includegraphics[scale=0.3]{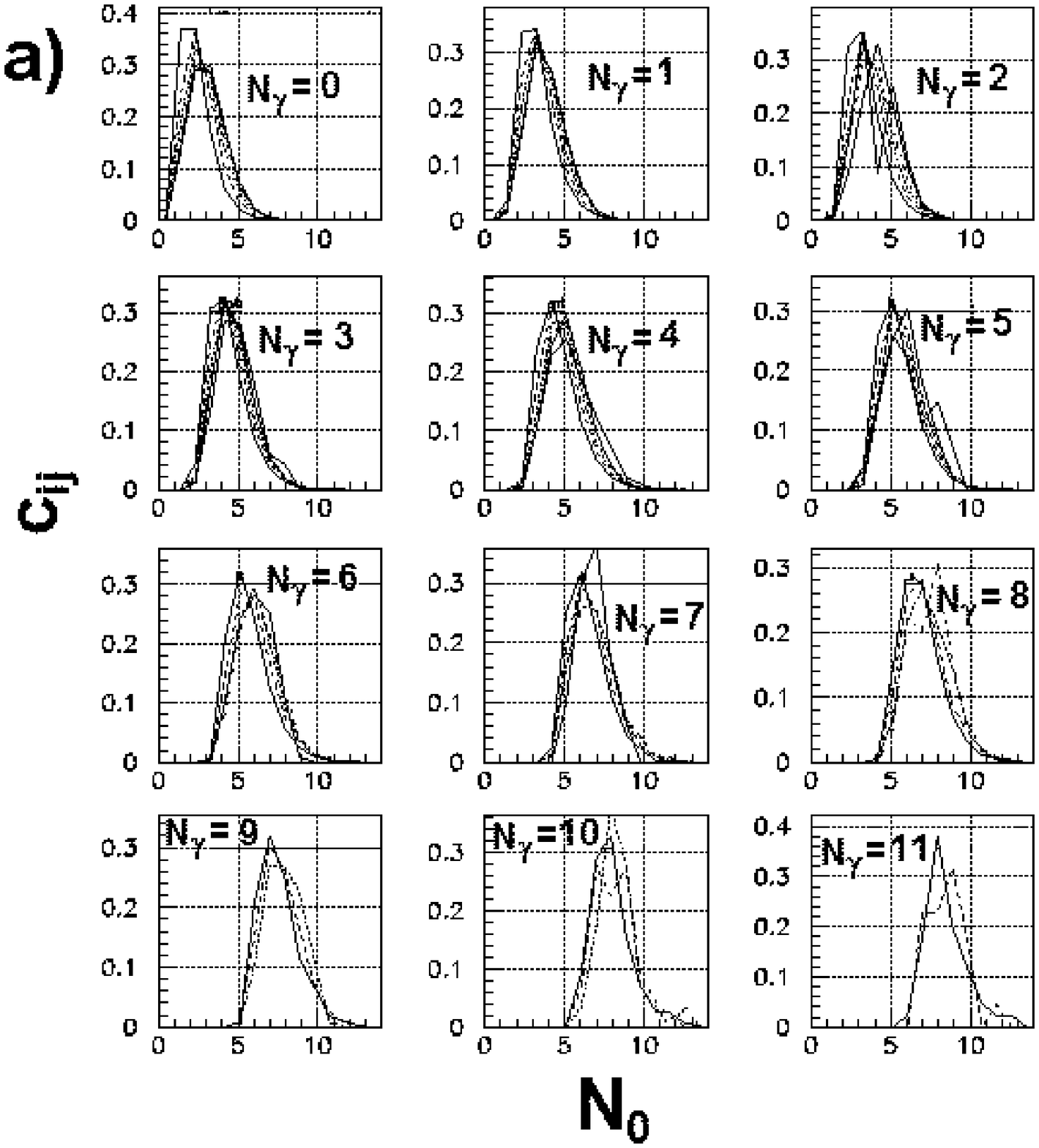}
\includegraphics[scale=0.3]{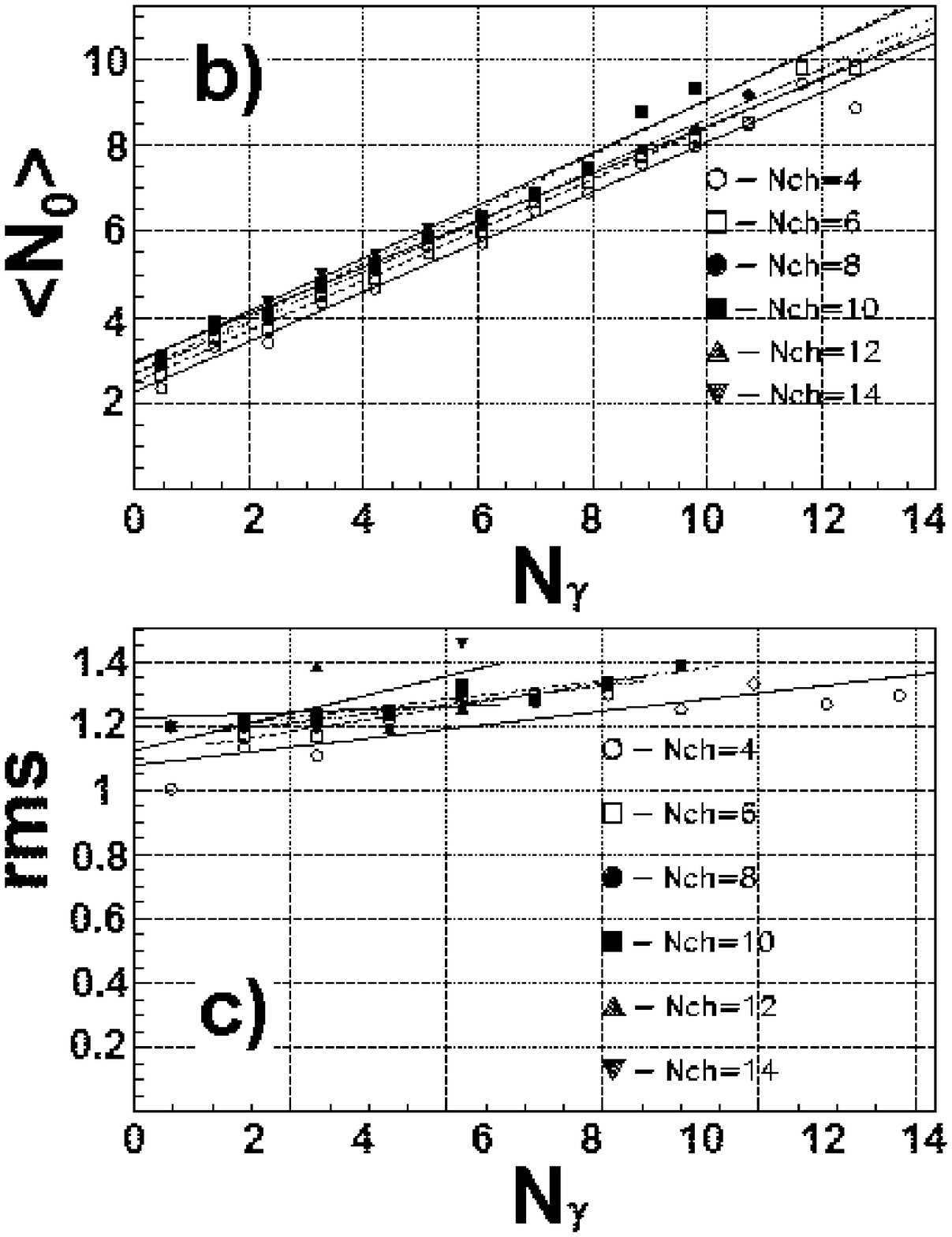}
\caption{\label{fig:wide}a) The dependence of $c_{ij}$ factors on $N_0$ for various $N_{\gamma}$ and $N_{ch}$, 
b) fitting parameters: $<N_0>$ and standard deviation (rms) as function of $N_{\gamma}$ for various $N_{ch}$.}
\end{figure*}

As mention before we have used scaled variables $n_0$ and $r_0$ (see Introduction): $n_0=N_0/N_{tot}$ and $r_0(n_0)=N_{ev}(N_0, N_{tot})/N_{ev}(N_{tot})$, where $N_{tot}=N_0+N_{ch}$. Function $r_0(n_0)$ 
are shown in Fig. 7 for every $N_{tot}>10$. All distributions are fitted with Gaussian function. The dependence of the fitting parameters is presented in Fig. 8. The data in the intervals (26, 27, 28) and (29, 30, 31) are combined due to small statistics.

\begin{figure*}
\includegraphics[scale=0.6]{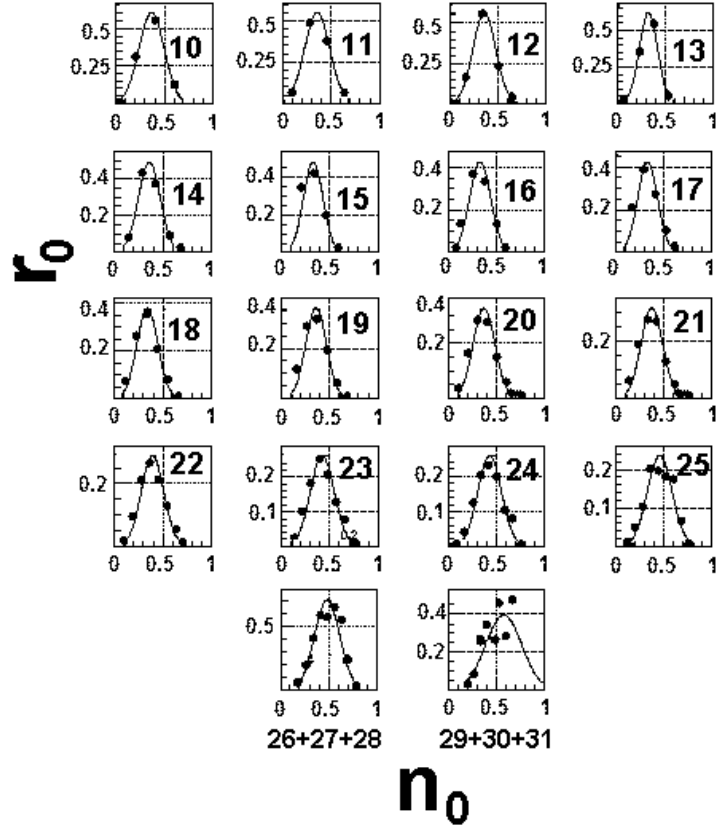}
\caption{\label{fig:wide}Scaled neutral pions number $n_0$ distributions for various $N_{tot}$ (are specified by number).}
\end{figure*}

One can see that the measured average $<n_0>$ (Fig. 8а) coincide with the same values for the neutral pions from MC events at $N_{tot}>$12. In the gluon dominance model (MGD) \cite{Kokoulina01} neutral pions average dependence on $N_{tot}$ has been received by analytical way. This dependence is also presented in Fig. 8а and illustrates quite good agreement with the experimental data at $N_{tot}>$14. The average $<n_{\gamma}>$ is also shown. The measured standard deviations, $\sigma$ (Gaussian), (Fig. 8b) have shown the qualitative agreement with MC model only for $N_{tot}<$22. The measured values $\sigma$ increase at higher $N_{tot}$.

\begin{figure*}
\includegraphics[scale=0.3]{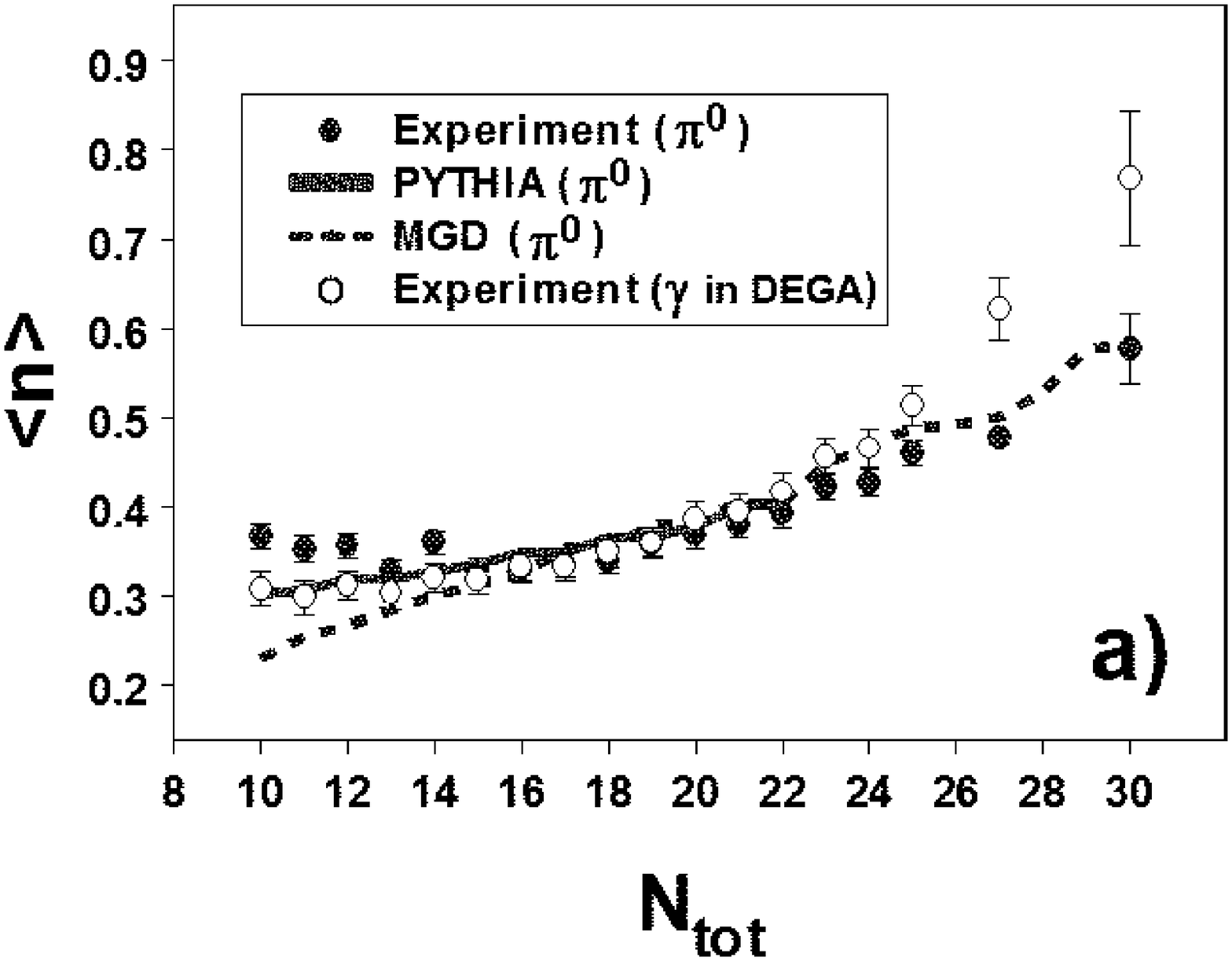}
\includegraphics[scale=0.3]{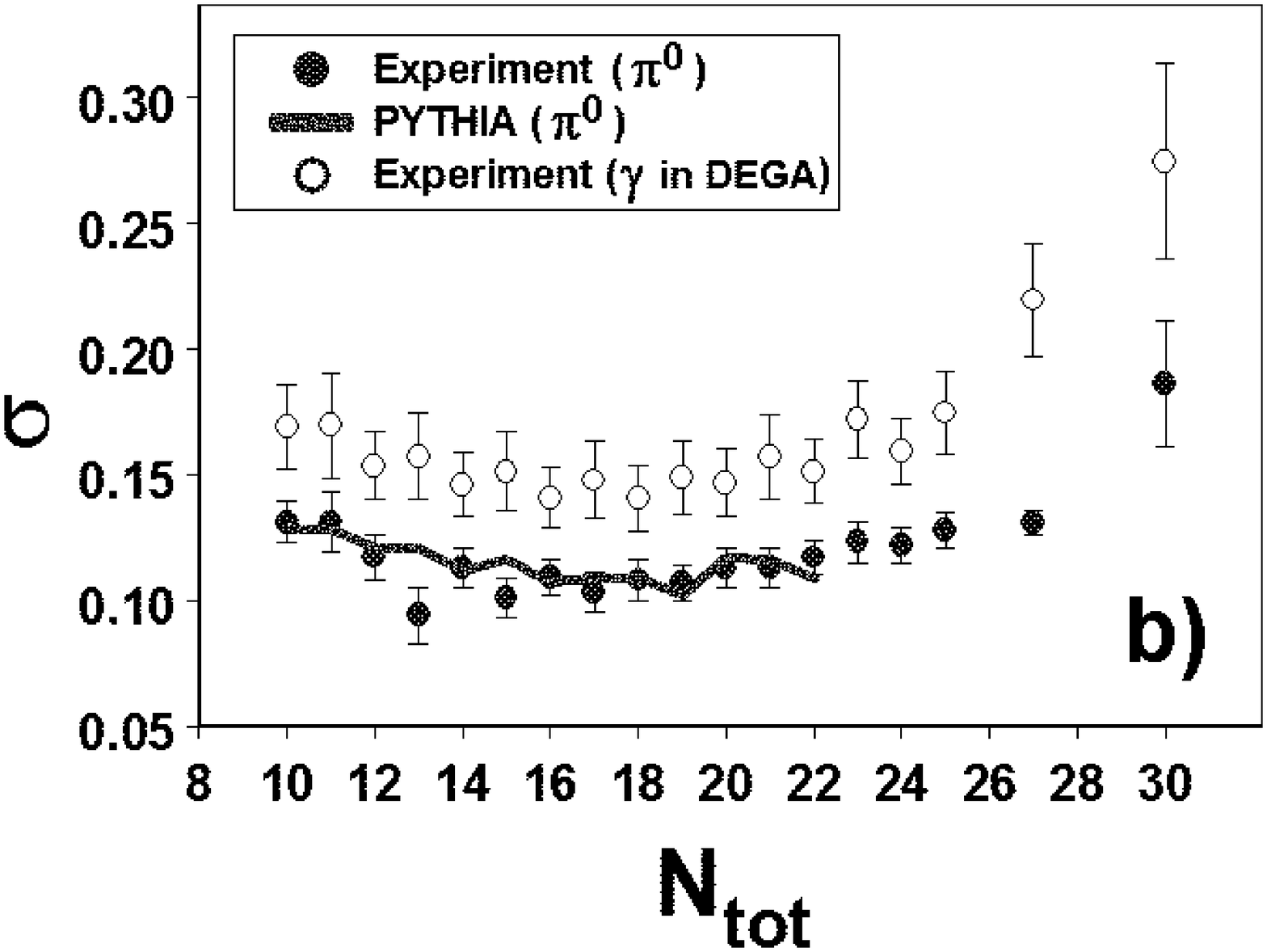}
\caption{\label{fig:wide}Fitted parameters of neutral pions number and photons number distributions for experimental data and МC events as function of $N_{tot}$.For neutral pions $N_{tot}=N_{ch}+N_0$, for photons $N_{tot}=N_{ch}+N_{\gamma}$.}
\end{figure*}

The theoretical prediction of scaled variance $\omega$ behavior (in our case $\omega=D(N_0)/<N_0>=\sigma^2\ast N_{tot}/<n_0>$) is given in \cite{Begun01}. The analysis has been done for three energy densities of the pion system at the approach to the Bose-Einstein condensate condition (pion condensate) (Fig. 9а). Our experimental data (Fig. 9b) have confirmed assumption on the BEC formation in pion system at $N_{tot}>$22 in pp-interactions at 50 GeV.

\begin{figure*}
\includegraphics[scale=0.35]{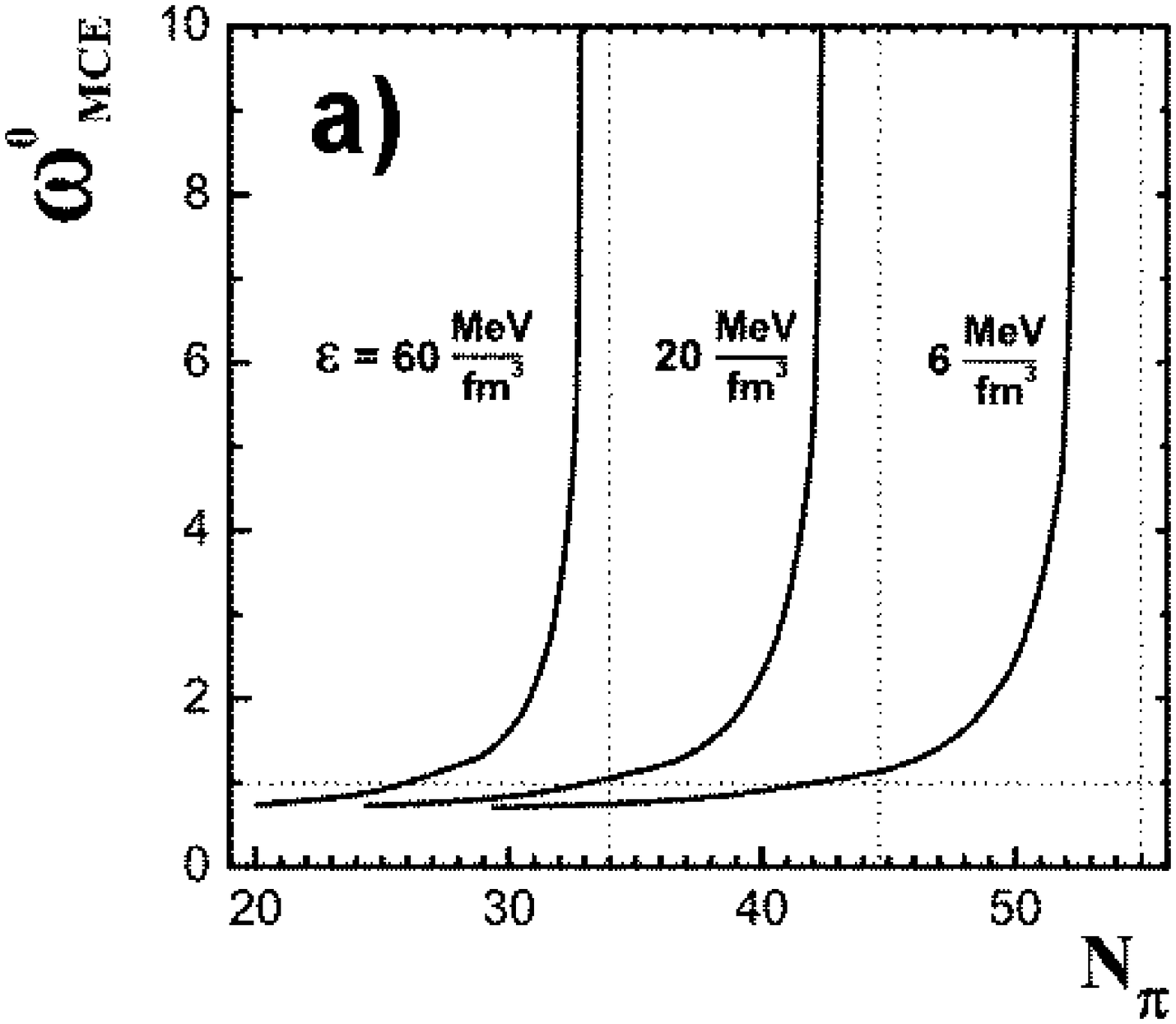}
\includegraphics[scale=0.35]{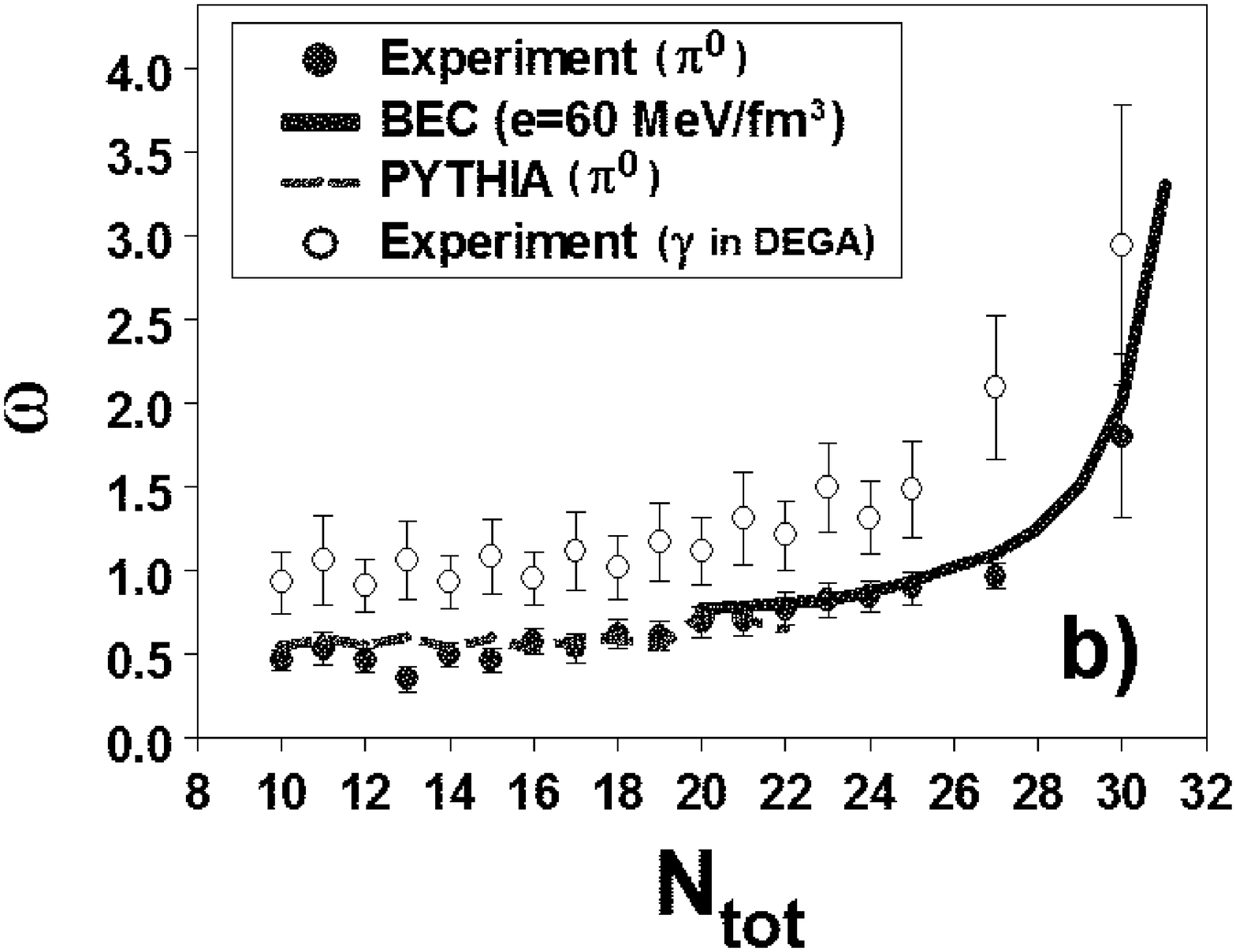}
\caption{\label{fig:wide}Scaled variance $\omega$ as function of $N_{tot}$ \cite{Begun01} and b) the result of the present measured of $\omega$ for neutral pions and photons. $N_{tot}=N_{ch}+N_0$ for neutral pions, $N_{tot}=N_{ch}+N_{\gamma}$ for photons.}
\end{figure*}

\section{Conclusion}
\label{conclusion}

Measurements of the charged and neutral pions number in the events with high multiplicity in pp-interactions at 50 GeV (experiment SERP-Е-190, SVD-2 setup) together with MC analysis led to the following results:
\begin{itemize}
\item The average number of neutral pions in the event is proportional to the photons number detected in DEGA calorimeter that allows one to extract pion number fluctuations from photon number fluctuations.
\item It is convenient to present data in the scaled form: $n_0=N_0/N_{tot}$ and $r_0(n_0)=N_{ev}(N_0, N_{tot})/N_{ev}(N_{tot})$ with interval $n_0$ is equal to 0$\div$1.
\item The corrections for limited aperture VD, trigger action and efficiency of data processing system have been introduced to the data.
\item The function $r_0(n_0)$ is fitted with Gaussian function and  values $<n_0>$, $\sigma$ and scaled variance $\omega=D/<N_0>$ are derived. They have shown the qualitative agreement with the same values received for PYTHIA5.6 code at $N_{tot}<$22.
\item Pion number fluctuations increase at $N_{tot}>$22, that indicates approaching to pion condensate conditions for the high multiplicity pion system according to GCE, CE, MCE models \cite{Begun01} \cite{Begun02}.
\item This effect has been observed for the first time.
\end{itemize}

This work was supported by the Russian Foundation for Basic Research (projects no. 08-02-90028  Bel-a, 09-02-92424  KE-a, 09-02-00445а, 06-02-16954) and was funded by a grant (no. 1456-2008-2) for support of leading scientific schools. 
	
Authors are grateful to a management of IHEP for support in carrying out of researches, to the staff of accelerator division and beam department for effective work of U-70 and the channel 22. Authors are appreciated to M. I. Gorenstein and V. V. Begun for stimulation of these studies and useful discussions.

\label{last}
\end{document}